\documentclass[11pt]{article}
\usepackage{graphicx,epsfig}

\newcommand{\BABARPubYear}    {05}

\newcommand{\BABARProcNumber} {056}
\newcommand{\SLACPubNumber} {11336}

\input babarsym
\def\Journal#1#2#3#4{{#1} {\bf #2}, #3 (#4)}

\def\PRL{\em Phys. Rev. Lett.}
\def\PRD{{\em Phys. Rev.} D}

\def\be{\begin{equation}}
\def\ee{\end{equation}}
\def\bea{\begin{eqnarray}}
\def\eea{\end{eqnarray}}

\def\thetaFive {\ensuremath{\Theta_5(1540)^+}\xspace}
\def\thetaFivenomass {\ensuremath{\Theta_5^+}\xspace}

\def\Xcc  {\ensuremath{X(3872)}\xspace}

\long\def\inst#1{\par\nobreak\kern 4pt\nobreak
    {\it #1}\par\vskip 10pt plus 3pt minus 3pt}

\begin{document}
{\pagestyle{empty}

\begin{flushright}
SLAC-PUB-\SLACPubNumber \\
\babar-PROC-\BABARPubYear/\BABARProcNumber \\
%\babar-PUB-\BABARPubYear/\BABARPubNumber \\
%hep-ex/\LANLNumber \\
July 6, 2005 \\
\end{flushright}

\par\vskip 4cm

% Title of the paper
\begin{center}
\Large \bf 
Pentaquarks and new hadron spectroscopy at \babar
\end{center}
\bigskip

\begin{center}
\large 
S. Ricciardi\\
Royal Holloway College, University of London \\
Department of Physics, Egham, Surrey TW20 0EX, UK \\
(for the \lbabar\ Collaboration)
\end{center}
\bigskip \bigskip

% Abstract
\begin{center}
\large \bf 
Abstract
\end{center}
Recent results on the search for pentaquarks
and on charmonium spectroscopy at \babar\ are reviewed.
The latter includes the observation of the puzzling new state
\Xcc\to\jpsi\pip\pim in \B decays, and the searches for
\Xcc in two-body \B decays and initial state radiation events.

\vfill
\begin{center}
Contributed to the Proceedings of the \\ XI$^{th}$ International 
Conference on Elastic and Diffractive Scattering, \\
5/15/2005---20/5/2005, Ch\^{a}teau de Blois, France
\end{center}

\vspace{1.0cm}
\begin{center}
{\em Stanford Linear Accelerator Center, Stanford University, 
Stanford, CA 94309} \\ \vspace{0.1cm}\hrule\vspace{0.1cm}
Work supported in part by Department of Energy contract DE-AC02-76SF00515.
\end{center}

\section{Introduction}
The \babar\ experiment is taking data at the PEP-II \epem collider
at the center of mass energy of 10.58 GeV.
The large data sample and the detector characteristics, 
which include excellent tracking and particle identification capabilities,
allow a rich and diversified hadronic 
physics programme. I will focus here on the results 
on possible candidates for new forms of matter, namely the
pentaquarks and the \Xcc, a charmonium-like resonance 
not fitting a pure \ccbar assignment.
Results are preliminary, unless otherwise specified.

\section{Pentaquarks}

Several experiments~\cite{pentareview} 
have recently claimed observation of
narrow baryonic resonances with exotic quantum numbers,
which are regarded as 5-quark candidates. 
The observed states are: \thetaFive (presumed structure $uudd\bar{s}$),
$\Xi_5(1860)^{--}$ ($ddss\bar{u}$) 
(and its non-exotic neutral partner $\Xi_5(1860)^{0}$),
and a charmed
resonance $\Theta_{5c}(3100)^{0}$ ($uudd\bar{c}$).
On the other hand, null results
have also been published and have now 
outnumbered the positive claims. 
The controversy between experimental evidences 
for and against the existence is fed by the problem of comparing
%measurements based on 
different production mechanisms and energy ranges.
Therefore, it is of interest to perform high statistics
and high resolution
searches which encompass different production processes,
as done by \babar.

\subsection{Inclusive search in \epem interactions}

We search for inclusive production of pentaquark states \epem \to $P\,X$
with any final state $X$ recoiling against the pentaquark $P$,
using 123 fb$^{-1}$ of data collected at the \epem center of mass (CM)
energy at or just below the mass of the \FourS resonance~\cite{babarpenta}.
In particular, we look for the lightest $P$ candidates:
\thetaFivenomass\to$p\KS$, 
$\Xi_5^{--}\to\Xi^{-}\pim$, and $\Xi_5^{0}\to\Xi^{-}\pip$.

The invariant mass of the $p\KS$ pair is shown in the left
plot in Fig.~\ref{fig:thetap}.
There is a clear
peak containing about 98,000 entries at 2285 \mevcc from $\Lambda_c^+\to p\KS$,
which demonstrates our detection sensitivity to narrow 
resonances. The inset shows a magnified view of the region where
the \thetaFive has been reported.
No enhancement is seen near 1540 \mevcc.
\begin{figure}
%\rule{5cm}{0.2mm}\hfill\rule{5cm}{0.2mm}
%\vskip 2.5cm
%\rule{5cm}{0.2mm}\hfill\rule{5cm}{0.2mm}
\center{
\epsfig{figure=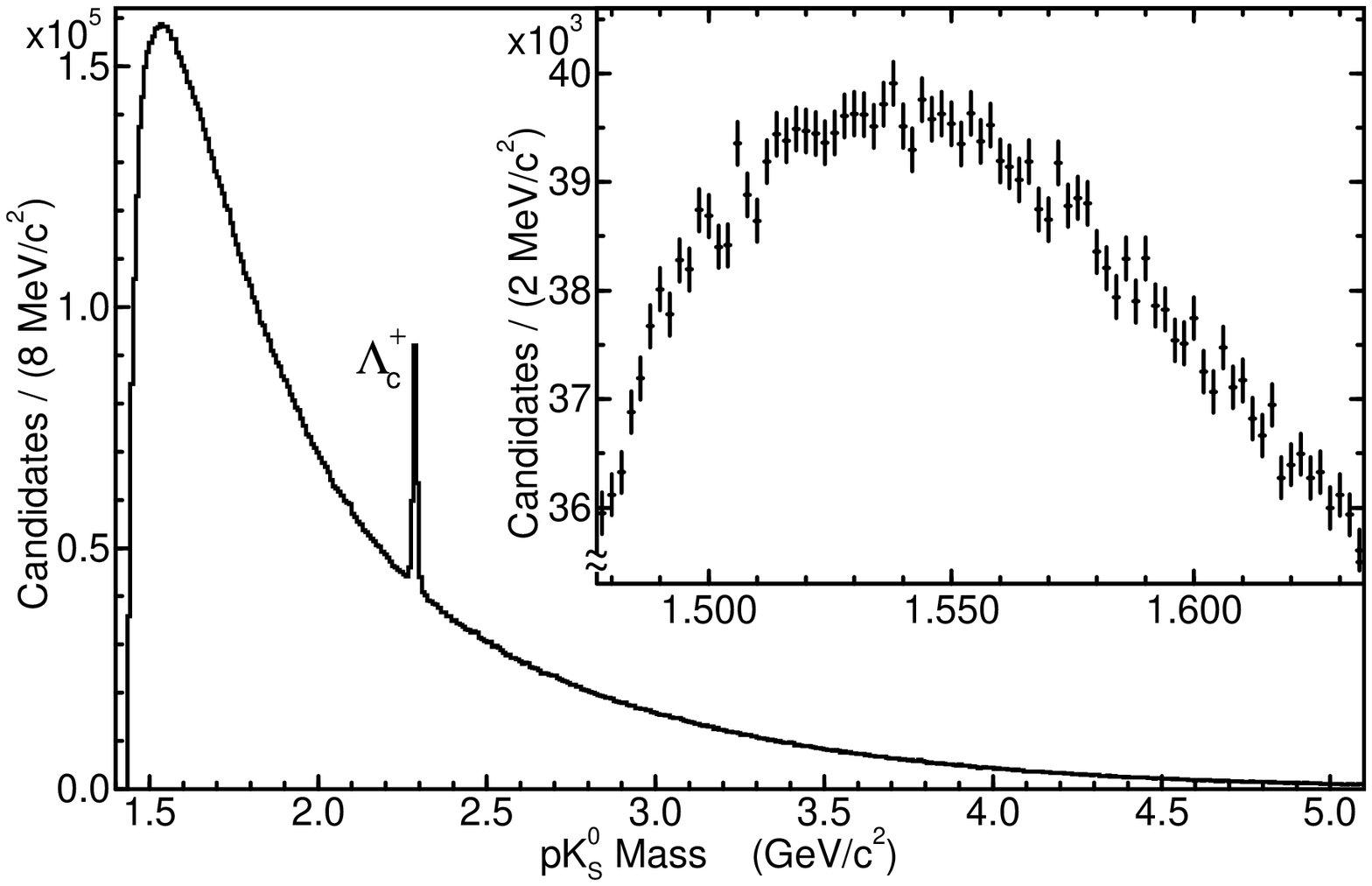,height=2.0in}
\epsfig{figure=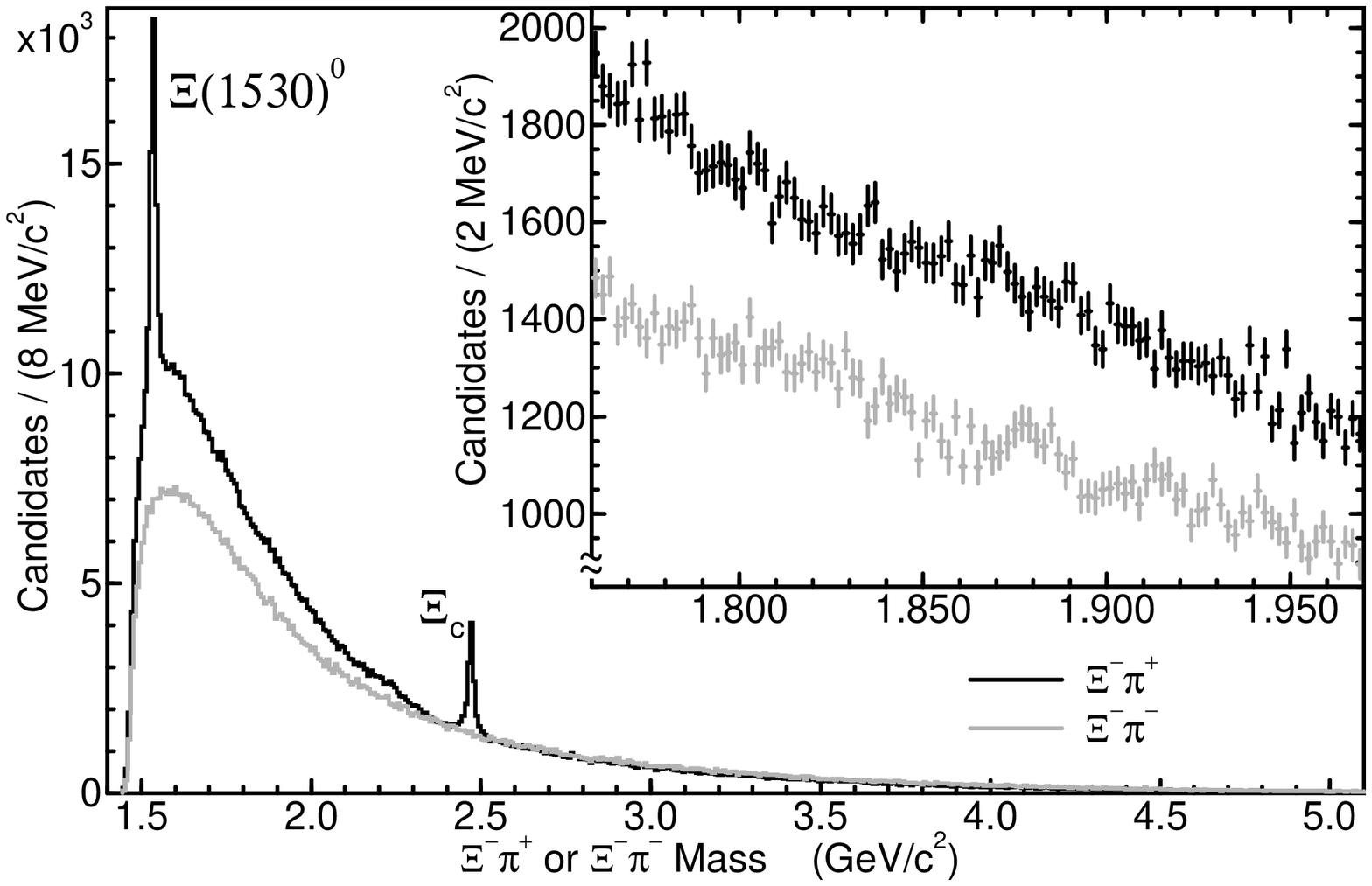,height=2.0in}}
\caption{Invariant mass distributions found in the
inclusive search for pentaquarks in \epem collisions:
$p\KS$ invariant mass (left) and of $\Xi^{-}\pi$ (right).
In the plot on the right distributions 
are shown for
$\Xi^{-}\pip$ (black) and $\Xi^{-}\pim$ (grey). The insets
are magnification of the mass regions in which pentaquark
states have been reported. 
}
\label{fig:thetap}
\end{figure}

To form the $\Xi_5^{--}$ and $\Xi_5^{0}$ candidates a $\Xi^{-}$
is combined with a like-sign or opposite-sign pion.
The $\Xi^{-}$ is reconstructed in its decay $\Xi^{-}\to\Lambda^{0}\pim$
with $\Lambda^{0}\to p\pim$. The invariant mass distributions
are shown in Fig.~\ref{fig:thetap}. The peaks
in the $\Xi^{-}\pip$ distribution are due to the
$\Xi^{0}(1530)$ and $\Xi^{0}_c(2470)$ baryons, but no structure
is visible where the $\Xi_5(1860)$ pentaquark is expected.

Null results are also obtained when the search is performed
separately in ten $p^*$ bins uniformly distributed between
zero and 5 \gevc, where $p^*$ is the momentum of the candidate in the
CM frame. 
A possible \thetaFive 
contribution is fitted in each $p^*$ bin
using two different hypotheses on the width, $\Gamma$: $\Gamma$ = 8 \mevcc,
corresponding to the experimental upper limit and larger than
our $p\KS$ resolution($\approx$ 2 \mevcc), and $\Gamma$ = 1 \mevcc.
A similar search is done for $\Xi_5^{--}$, assuming as widths the upper
limit $\Gamma$ = 18 \mevcc, and $\Gamma$ = 1 \mevcc.

These null results are used to set limits on the
differential production cross section of \thetaFive 
and $\Xi_5^{--}$ in \epem interactions,
shown in Fig.~\ref{fig:differential}, assuming
${\cal{B}}(\thetaFive\to p\KS) = 1/4$
and ${\cal{B}}(\Xi_5^{--}\to\Xi^-\pim) = 1/2$. 
\begin{figure}
\center{
\epsfig{figure=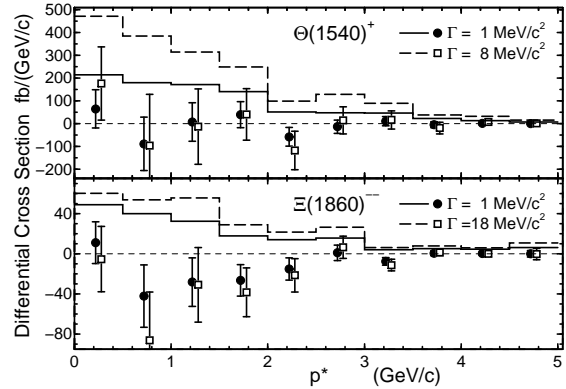,height=2.0in}}
\caption{
The measured differential production cross sections
(symbols) and corresponding 95\% CL upper limits (lines)
for \thetaFivenomass (top) and $\Xi_5^{--}$ (bottom),
assuming natural widths of $\Gamma$ = 1 \mevcc (solid) and at the current 
experimental upper limit (open/dashed), as function of CM momentum.
}
\label{fig:differential}
\end{figure}
The corresponding upper limits at 95\% CL on the total production rates 
per $q\bar{q}$ event
are 11$\times 10^{-5}$ and 1.1$\times 10^{-5}$, respectively, for the maximum widths.
These values are about a factor eight 
and four below the typical values measured for ordinary
baryons of the same mass.

\subsection{Search of \thetaFive in electro- and hadro-production}

In addition to \epem collisions, we have analysed
events due to the interactions with the detector material 
of beam-halo electrons and positrons (electro-production), 
and of hadrons from the
primary collision (hadro-production).
The target material is mainly due to the Be beam-pipe and
to the inner detector (a five layers silicon-vertex tracker).
We search for \thetaFive\to$p\KS$ using simple cuts on the
position of the vertex, on the invariant mass of the \KS, and
\dedx information to identify protons and charged pions. 
The spatial distribution of the $p\KS$ vertices
reproduces to high accuracy the known geometry of the beam-pipe
and of the inner detector, giving confidence that
these events are due to interactions in the detector material.  
The momentum distribution of the \KS and proton
are almost entirely confined to the region below 1.5\gevc, thus 
very similar to those measured by many experiments which have
observed \thetaFive. However, our 
$p\KS$ invariant mass distribution shows no evidence of a structure
near 1.54 \gevcc in a sample corresponding to 230~fb$^{-1}$ of data.

\section{Results on the \Xcc}

The discovery of the \Xcc\to\jpsi\pip\pim by the Belle Collaboration~\cite{belleX}
has been confirmed by CDF~\cite{cdfX}, D0~\cite{d0X} and \babar~\cite{babarX}.
While the existence of this high mass narrow resonance is not questioned, its
nature is still very much debated. The most likely charmonium 
candidate is the $J^{PC} =2^{--}$, $^3D_2$ $\psi_2$ state, which, however, should 
have a large radiative transition into $\chi_c$,
which was not observed~\cite{belleX}. 
Inspired by 
its mass, right at the \Dz\Dstarzb threshold,
a number of non-Standard Model interpretations have recently been proposed
including a \Dz\Dstarzb  molecule model
and diquark-antidiquark model.
In the molecule model~\cite{moleculeX} the \Xcc is a loosely bound S-wave \Dz\Dstarzb,
with quantum numbers $J^{PC} =1^{++}$, that is produced in weak decays
of \B mesons into \Dz\Dstarzb\kaon. Using factorization, heavy-quark
symmetry and isospin symmetry, the branching fraction
for $B^0\to\Xcc K^0$ is predicted to be suppressed by more than an order
of magnitude compared to that for $B^+\to\Xcc K^+$.
The diquark-antidiquark model~\cite{diquarkX} predicts two neutral states
in the mass range of the \Xcc, with different quark composition, 
$X_u$ = [$\bar{c}\bar{u}$][cu] and $X_d$ = [$\bar{c}\bar{d}$][cd],
which can mix.
If one amplitude (from $X_d$ or $X_u$) is dominant in the charged
mode and the other in the neutral mode, the model predicts the rates
to be equal and the mass difference to be ($7 \pm 2$)\mevcc.
This model also predicts the existence of charged partners of the \Xcc,
at a level which is not excluded by a previous \babar\ analysis~\cite{babarXch}.

We reconstruct the exclusive decay 
$\Bm\to \Xcc K^{-}$
and
$\Bz\to \Xcc K^{0}_S$, $\Xcc\to\jpsi\pip\pim$, in a data sample
corresponding to the integrated luminosity of 211 fb$^{-1}$.
The distributions of the invariant mass $m_X$ of the $\jpsi\pip\pim$
system are shown in Fig.~\ref{fig:x3872}.
\begin{figure}
\center{
\epsfig{figure=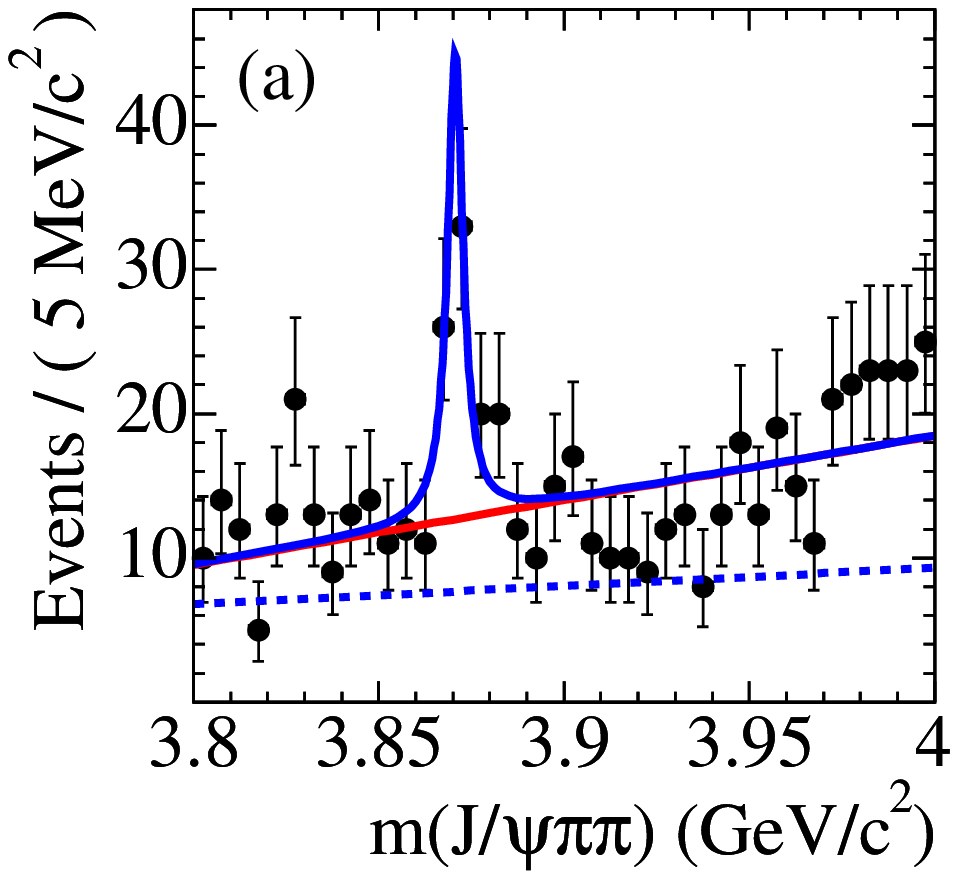,height=2.0in}
\epsfig{figure=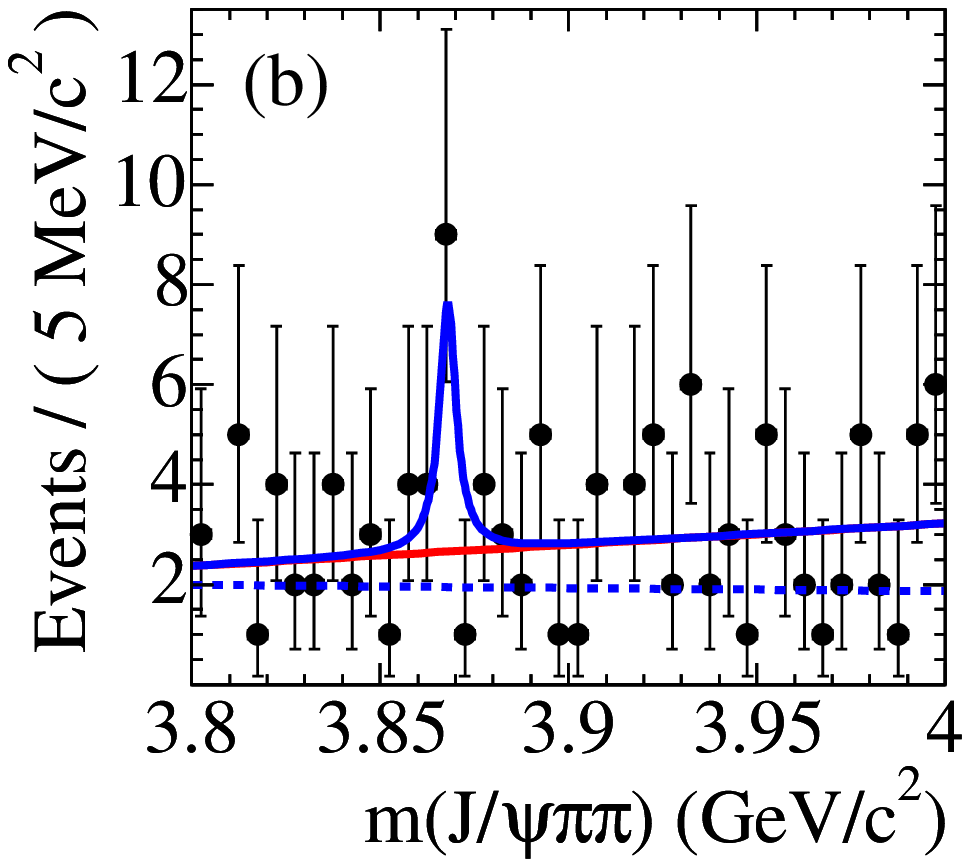,height=2.0in}}
\caption{Invariant mass distributions $m_X$ for
(a) $\Bm\to \Xcc K^{-}$ and (b) $\Bz\to \Xcc K^{0}_{S}$.
The lines represent the result of the projection on $m_X$
of the fit used to extract the yields and the resonance parameters. 
A clear signal peak is visible over the linear background components.
}
\label{fig:x3872}
\end{figure}
We find a 5.7$\sigma$ significance for the \Bm mode and 2.5$\sigma$
for the \Bz mode, including systematic uncertainties.
The mass difference of the \Xcc produced in \Bz and \Bm decays is
$\Delta m =(2.7 \pm 1.3\pm 0.2)$ \mevcc
and is both consistent with zero and with the diquark-antidiquark 
model prediction.
From the number of signal events, the efficiencies, the secondary 
branching fractions and the number of \BB events, we obtain the branching fraction products
${\cal B}(\Bm\to \Xcc K^{-}, \Xcc \to \jpsi \pip\pim) = (8.5 \pm 2.4 \pm 0.8) \times 10^{-6}$
and
${\cal B}(\Bz\to \Xcc K^{0}_S, \Xcc \to \jpsi \pip\pim) =(5.1 \pm 2.8 \pm 0.7) \times 10^{-6}$.
The ratio of neutral to charged branching fractions $R = 0.61\pm 0.36\pm0.06$
is more consistent with isospin conserving decays ($R = 1$), than the 
molecule model ($R < 0.1$).

In addition, we have used a complementary approach, based on the measurement
of the kaon momentum spectrum in the \B rest frame, 
to measure the inclusive \B decays to two-body final state
$\B\to\kaon X$, where $X$ can be any state, including the \Xcc.
Two body decays can be identified by their characteristic
monochromatic line, and no reconstruction of the $X$
decay is necessary. Therefore, this method allows 
to determine the absolute branching fraction (or set
upper limits) for production of all known charmonium
resonances and the \Xcc. The analysis is performed
on a sample of \FourS events where a candidate \B meson 
is fully reconstructed, so that the momentum of the recoiling
\B can be calculated from the measured \B and the beam
parameters. We find a significant signal for \jpsi, \etac,
\chicone, \psitwos, but no evidence of \Xcc.
We derive ${\cal B}(\Bm\to \Xcc K^{-})<3.2 \times 10^{-4}$,
which, in conjunction with the branching fraction
product, allows to set the lower limit
${\cal B}(\Xcc\to\jpsi\pip\pim) > 4.2\%$ at 90\% C.L.

In order to narrow down the quantum numbers of the \Xcc,
we have also looked for \Xcc\to\jpsi\pip\pim
in \epem initial state radiation events, allowed for $J^{PC}=1^{--}$.
We find no evidence of a signal and set an upper limit
on the product ${\cal B}(\Xcc\to\jpsi\pip\pim)\times\Gamma_{ee}^X < 6.2$ eV 
~\cite{babarISR} at 90\% C.L.

\section{Conclusions}

In summary, \babar\ is carrying out several sensitive searches
for pentaquarks and new hadrons. 
Pentaquark production has given null results in both \epem interactions
and fixed-target electro- and hadro-production. The upper limits
on the production rates in \epem are well below the rates for 
ordinary baryons of similar masses. The energy range probed
in the fixed-target experiment is similar to that of
previous electro-production experiments that have observed \thetaFive,
suggesting that prior claims are less than convincing.

The narrow state \Xcc has been confirmed by \babar\ in the decay
of \B mesons. Measurements of the mass and branching fractions 
have been presented for the first time separately for charged and neutral \B decays.
As the statistical precision will increase they
will allow to distinguish among different hypotheses on the \Xcc nature. 

%\section*{Acknowledgments}

%\section*{Appendix}

%\section*{References}

\end{document}